\documentclass[
 reprint,
superscriptaddress,
prb,
amsmath,amssymb,
 aps,
 longbibliography,
floatfix,
]{revtex4-1}

\usepackage{graphicx}
\usepackage{dcolumn}
\usepackage{bm}
\usepackage{array}
\usepackage{mathtools}
\usepackage{amsmath}
\newcolumntype{L}{>{\centering\arraybackslash}m{3cm}}
\usepackage[mathlines]{lineno}
\usepackage{times}
\usepackage{fouriernc}
\usepackage{bookmark}
\usepackage{comment}
\usepackage[colorinlistoftodos]{todonotes}
\usepackage{hyperref}
\PassOptionsToPackage{unicode}{hyperref}
\PassOptionsToPackage{naturalnames}{hyperref}

\newcommand{\Agx}{CeCu$_{6-x}$Ag$_{x}$}

\newcommand{\Aux}{CeCu$_{6-x}$Au$_{x}$}
\newcommand{\Agq}{CeCu$_{5.8}$Ag$_{0.2}$}

\newcommand{\Auq}{CeCu$_{5.9}$Au$_{0.1}$}
\newcommand{\Ag}{CeCu$_{5.8}$Ag$_{0.2}$}
\newcommand{\bfQ}{{\bf Q}}
\newcommand{\bfq}{{\bf q}}

\usepackage{cleveref}

\crefname{equation}{equation}{equations}
\Crefname{equation}{Equation}{Equations}
\crefrangelabelformat{equation}{(#3#1#4--#5#2#6)}

\crefmultiformat{equation}{equations (#2#1#3}{, #2#1#3)}{#2#1#3}{#2#1#3}
\Crefmultiformat{equation}{Equations (#2#1#3}{, #2#1#3)}{#2#1#3}{#2#1#3}

\begin{document}

\preprint{APS/123-QED}

\title{Unveiling the role of competing fluctuations at an unconventional quantum critical point}

\author{L. Poudel}
\email{lpoudel@vols.utk.edu}
\affiliation{Department of Physics \& Astronomy, University of Tennessee, Knoxville, TN-37996, USA}
\affiliation{Quantum Condensed Matter Division, Oak Ridge National Laboratory, Oak Ridge, TN-37831, USA}
\affiliation{Department of Materials Science \& Engineering, University of Maryland, College Park, MD 20742}
\affiliation{NIST Center for Neutron Research, Gaithersburg, MD-20899}

\author{J. M. Lawrence}
\affiliation{Los Alamos National Laboratory, Los Alamos, NM-87545, USA}

\author{L. S. Wu}
\affiliation{Quantum Condensed Matter Division, Oak Ridge National Laboratory, Oak Ridge, TN-37831, USA}

\author{G. Ehlers}
\affiliation{Quantum Condensed Matter Division, Oak Ridge National Laboratory, Oak Ridge, TN-37831, USA}

\author{Y. Qiu}
\affiliation{NIST Center for Neutron Research, Gaithersburg, MD-20899}

\author{A. F. May}
\affiliation{Materials Science \& Technology Division, Oak Ridge National Laboratory, Oak Ridge, TN-37831, USA}

\author{F. Ronning}
\affiliation{Los Alamos National Laboratory, Los Alamos, NM-87545, USA}

\author{M. D. Lumsden}
\affiliation{Quantum Condensed Matter Division, Oak Ridge National Laboratory, Oak Ridge, TN-37831, USA}

\author{D. Mandrus}
\affiliation{Department of Physics \& Astronomy, University of Tennessee, Knoxville, TN-37996, USA}
\affiliation{Materials Science \& Technology Division, Oak Ridge National Laboratory, Oak Ridge, TN-37831, USA}
\affiliation{Department of Material Science \& Engineering, University of Tennessee, Knoxville, TN-37996, USA}

\author{A.D. Christianson}
\email{christiansad@ornl.gov}

\affiliation{Quantum Condensed Matter Division, Oak Ridge National Laboratory, Oak Ridge, TN-37831, USA}
\affiliation{Department of Physics \& Astronomy, University of Tennessee, Knoxville, TN-37996, USA}

\date{\today}


\begin{abstract}

Quantum critical points (QCPs) are widely accepted as a source of a diverse set of collective quantum phases of matter. A central question is how the order parameters of phases near a QCP interact and determine the fundamental character of the critical dynamics which drive the quantum critical behavior. One of the most interesting proposals for the quantum critical behavior that occurs in correlated electron systems is that the behavior may arise from local, as opposed to long wavelength, critical fluctuations of the order parameter. The local criticality is believed to give rise to energy over temperature ($E/T$) scaling of the dynamic susceptibility with a fractional exponent near the quantum critical point (QCP). Here we show that $E/T$ scaling is indeed observed for CeCu$_{6-x}$Ag$_x$ but on closer inspection, the fluctuations can be separated into two components, implying that multiple order parameters play an important role in the unconventional critical behavior. Additionally, when the fluctuations corresponding to the magnetically ordered side of the phase diagram are separated, they are found to be three dimensional and to obey the scaling behavior expected for long wavelength fluctuations near an itinerant antiferromagnetic QCP.
\end{abstract}

\maketitle

The QCPs found in heavy fermion materials serve as prototypes of quantum criticality\cite{Loh_REV,Gegenwart2008,Si1161}. However, several fundamental characteristics remain unexplained preventing a general understanding of quantum critical phenomena. For example, in several heavy fermion QCPs, heat capacity measurements suggest that the effective mass of the quasiparticles diverges logarithmically\cite{Loh_REV,custers2003break,Gegenwart2003,UCu4Pd,CeCoIn5,Lohneysen1996471}. Furthermore, in UCu$_{5-x}$Pd$_x$ and \Aux, energy over temperature ($E/T$) scaling of the critical dynamics demonstrates the equivalency of $E$ and temperature $T$ at a QCP. Beyond these cases, $E/T$-scaling provides a stringent test of theories across a spectrum of quantum materials including highly frustrated magnets, spin liquids, high $\mathrm{T_c}$ superconductors, non-Fermi liquid behavior, and quantum phase transitions \cite{aeppli1997nearly,Nicklas99,spin_liquid,PhysRevLett.75.725,lake2005quantum}. In \Aux, $E/T$ scaling is found with an unusual fractional exponent, which also appears in the temperature dependence of magnetic susceptibility \cite{schroder2000onset,schroder1998scaling}, indicating that the QCP is much different from the conventional framework proposed by Hertz, Millis, and, Moriya (HMM) \cite{hertz1976quantum, MillisPRB, moriya1985spin}. Recently, the debate as to the understanding of these phenomena has intensified \cite{Si_local,si2001locally,Friedemann17082010,strong_coupling,strong_coupling_PNAS, dissi_xy,dissipation_Varma,coleman2001}. This renewed debate is particularly significant since the hallmark of the unconventional quantum criticality, the unusual $E/T$ scaling, is explained by disparate mechanisms such as the breakdown of a local energy scale \cite{Si_local,si2001locally,Friedemann17082010}, coupling between quasiparticles and order parameter fluctuations \cite{strong_coupling,strong_coupling_PNAS}, or by topological excitations \cite{dissi_xy,dissipation_Varma}. As $E/T$ scaling of the dynamic susceptibility is almost universally used to validate theories of quantum criticality, additional more stringent experimental tests are essential.

 Here, we approach the aforementioned questions by studying the spin dynamics of the QCP composition \Agq{}. The compound series \Agx{} is a member of a family of materials based on the heavy fermion compound CeCu$_6$ \cite{Loh_REV}, which serves as an important prototype of unconventional QCPs\cite{Loh_REV,poudel}. At larger values of $x$, long ranged antiferromagnetic order is observed in \Agx{}, which is characterized by amplitude modulated $Ce-$moments with an incommensurate wave-vector $\bfQ$ = (0.65 0 0.3) \cite{poudel,Gango_long}. At the QCP, \Agq{} displays a logarithmic divergence of the heat capacity over temperature (C/T) suggesting physics beyond the HMM model \cite{Kuchler2004}. Additionally, the divergence of the Gr\"uneisen ratio is much weaker than the predictions of HMM model \cite{Kuchler2004}. Hence, \Agq{} provides an ideal model system to investigate the microscopic origins of the quantum critical behavior. 

In this paper, we present a detailed investigation of the quantum critical behavior of \Agq{} using state of art inelastic neutron scattering (INS) spectroscopy. Scaling analysis of the imaginary part of the dynamic susceptibility, $\chi"$ as a function of $T$ and energy transfer, $E$ is performed over a large region of reciprocal space. $\chi"$ at various momentum transfers ($\bfQ$s) is described by a single phenomenological equation which yields an unusual scaling exponent, $\alpha = 0.73(1)$ demonstrating unconventional QCP behavior. Further analysis of the fluctuation spectrum reveals the presence of fluctuations at multiple reciprocal space positions. Once separated, the fluctuations occurring at reciprocal space positions corresponding to magnetic order are found to be three dimensional and can be scaled according to the predictions of HMM theory. This suggests that the interplay of multiple order parameters play an important role in the quantum critical behavior in the class of QCPs exemplified by \Aux~and \Agx.
\\

\begin{figure}
\includegraphics[width=0.45\textwidth]{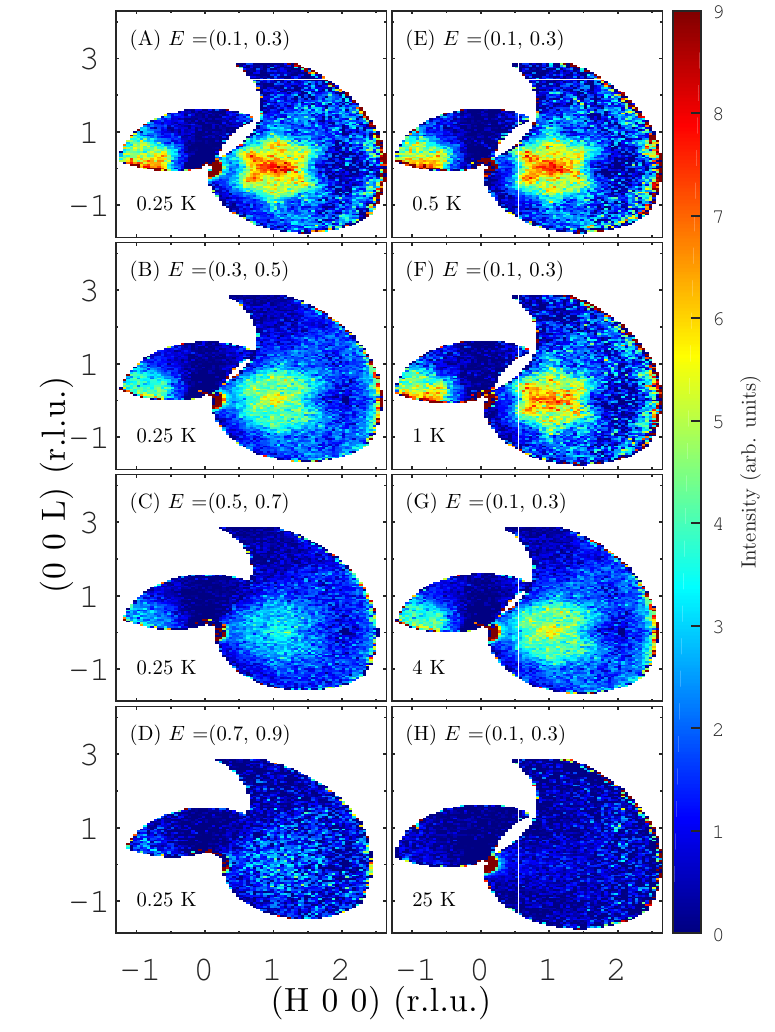}
\caption{Constant energy slices of the INS data from \Agq{} collected with CNCS. (A-D) The $E$ dependence of the scattering in (H 0 L) scattering plane at 0.25 K. (E-H) The $T$ dependence of the magnetic scattering in the interval $E$ = (0.1, 0.3) meV. In all panels, data collected at 50 K is used as the background and is subtracted from the data. Each pixel in the plot represents the integrated intensity in the area of dimension (0.05)$^2$ r.l.u.$^2$. r.l.u. denotes reduced lattice units. }
\label{slices}
\end{figure}

\begin{figure}
\centering
\includegraphics[clip, trim=0inch 0inch 0.0inch 0inch, width= 0.48\textwidth]{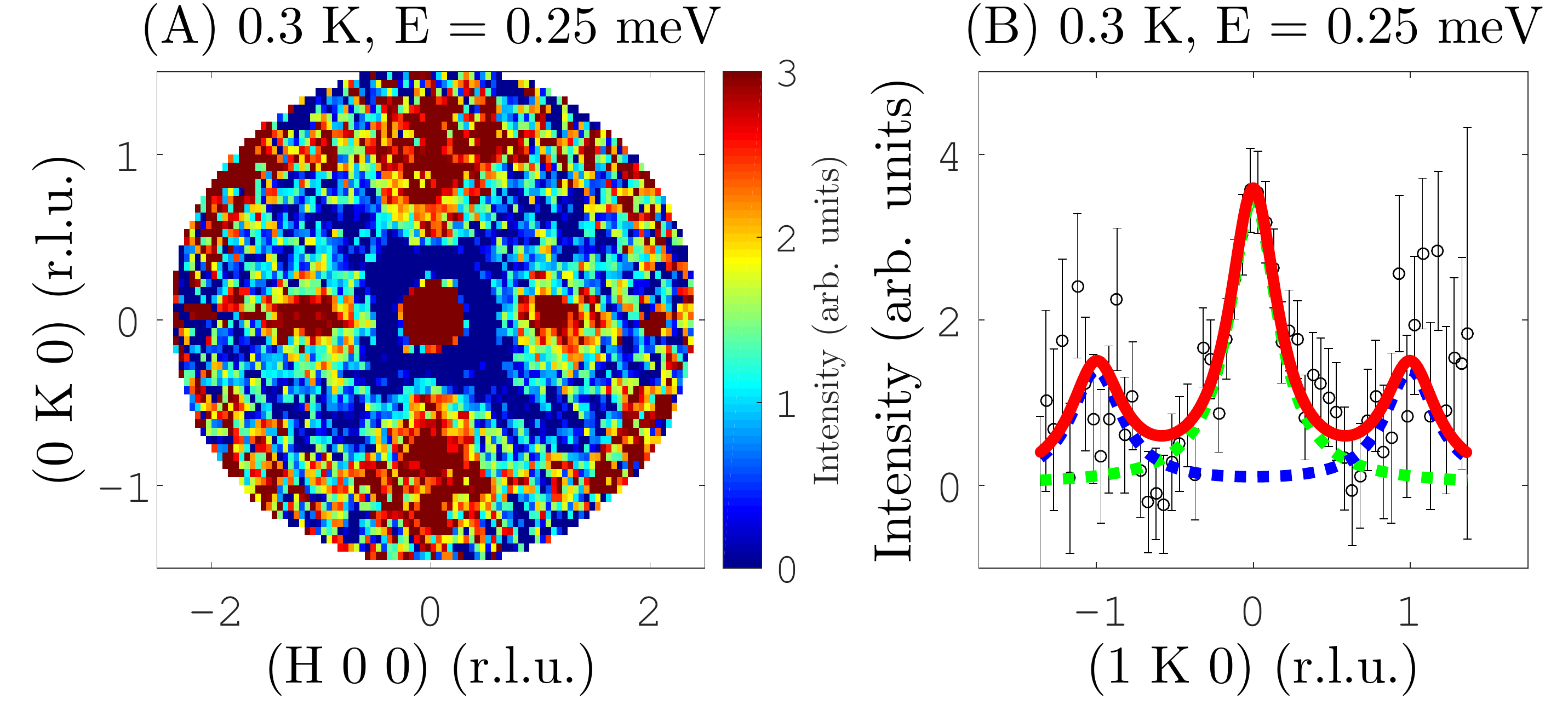}
\caption{Inelastic neutron scattering data collected with MACS at 0.3 K with 50 K data subtracted to remove background. (A) Constant $E$ slice at 0.25 meV showing the magnetic scattering in the (H K 0) scattering plane. Each pixel in the plot represents the integrated intensity in the area of dimension $\mathrm{0.05^2~r.l.u.^2}$. (B) Cut along (1 K 0). The green and blue dotted lines represent fits of the Lorentzian function to the fluctuations centered at $\bfQ$ = $\bfQ$ = (1 0 0) and (1 $\pm$1 0), respectively. The red line is the sum of Lorentzian components. }
\label{HK0}
\end{figure}

\hspace{-12pt}
\textbf{Results}\\
\textbf{Magnetic Scattering.} Sample preparation and the inelastic neutron scattering (INS) measurements are described in the \textbf{Methods} section and additional details are given in the supplementary information (SI) \cite{supp}. Several methods were used to extract the magnetic signal from the measured spectra, the details of which are described in the SI \cite{supp}. The results obtained from all methods are similar, providing confidence in the determination of magnetic scattering and subsequent analysis. In addition, thermodynamic measurements show that heat capacity over temperature ($C/T$) logarithmically diverges with temperature, indicating that the sample is close to the QCP (Fig. S2).

Figure \ref{slices} shows the evolution of the magnetic scattering in the (H 0 L) scattering plane with $E$ and $T$. Magnetic scattering is observed in the form of a diffuse pattern centered at $\bfQ$=(100). The intensity of the scattering is maximum at (100) but does not strongly vary in the nearby region of reciprocal space. With increasing $E$, the intensity of the scattering decreases and becomes more diffuse as shown in Figs. \ref{slices}(A-D). A qualitatively similar situation arises with increasing $T$. As shown in Figs. \ref{slices}(A,E-H), the pattern of scattering remains essentially unchanged until 1 K but becomes more diffuse at higher temperatures. A more detailed analysis is presented in the following sections.

Figure \ref{HK0} shows the scattering in the (H K 0) scattering plane. In addition to the magnetic scattering described above, scattering centered at (0 1 0) (Fig. \ref{HK0}(A)) is observed. The correlation length along the $b$-axis was determined with a fit of a Lorentzian function to a cut along (1 K 0), which is shown in Fig. \ref{HK0}(B). The fit yields $\xi_b$ = 28(1) $\mbox{\AA}$ which, as discussed in more detail below, is comparable to the real space correlation lengths in the $ac-$plane.\\

\hspace{-12pt}
\textbf{\textit{E/T}-scaling.}
For a more quantitative analysis of the magnetic scattering, we use the same phenomenology used previously to examine the critical dynamics of \Auq{} \cite{schroder1998scaling,schroder2000onset}. This phenomenology relates the magnetic susceptibility with $E$, $T$, and \bfQ{} and is rooted in the Curie-Weiss law \cite{schroder2000onset}. Within this picture, the dynamic susceptibility, $\chi(\bfQ,E,T)$ can be written as,
\begin{equation}
 \chi(\bfQ,E, T) =\frac{C}{\theta^{\alpha}+(T-iE)^{\alpha}}
\label{scaling}
 \end{equation}
 where $\theta(\bfQ-\bfQ_0)$ captures the wave-vector dependence of the magnetic fluctuations similar to the Curie-Weiss temperature. $\alpha$ is a scaling exponent independent of $\bfQ$, $T$, and $E$. The mean field limit is given by $\alpha$ = 1. This equation yields $E/T$ scaling at the critical wave-vector and $E/\theta(\bfq)$- scaling at zero temperature \cite{schroder2000onset}. 
 
 \begin{figure}
\centering
\includegraphics[width=0.48\textwidth]{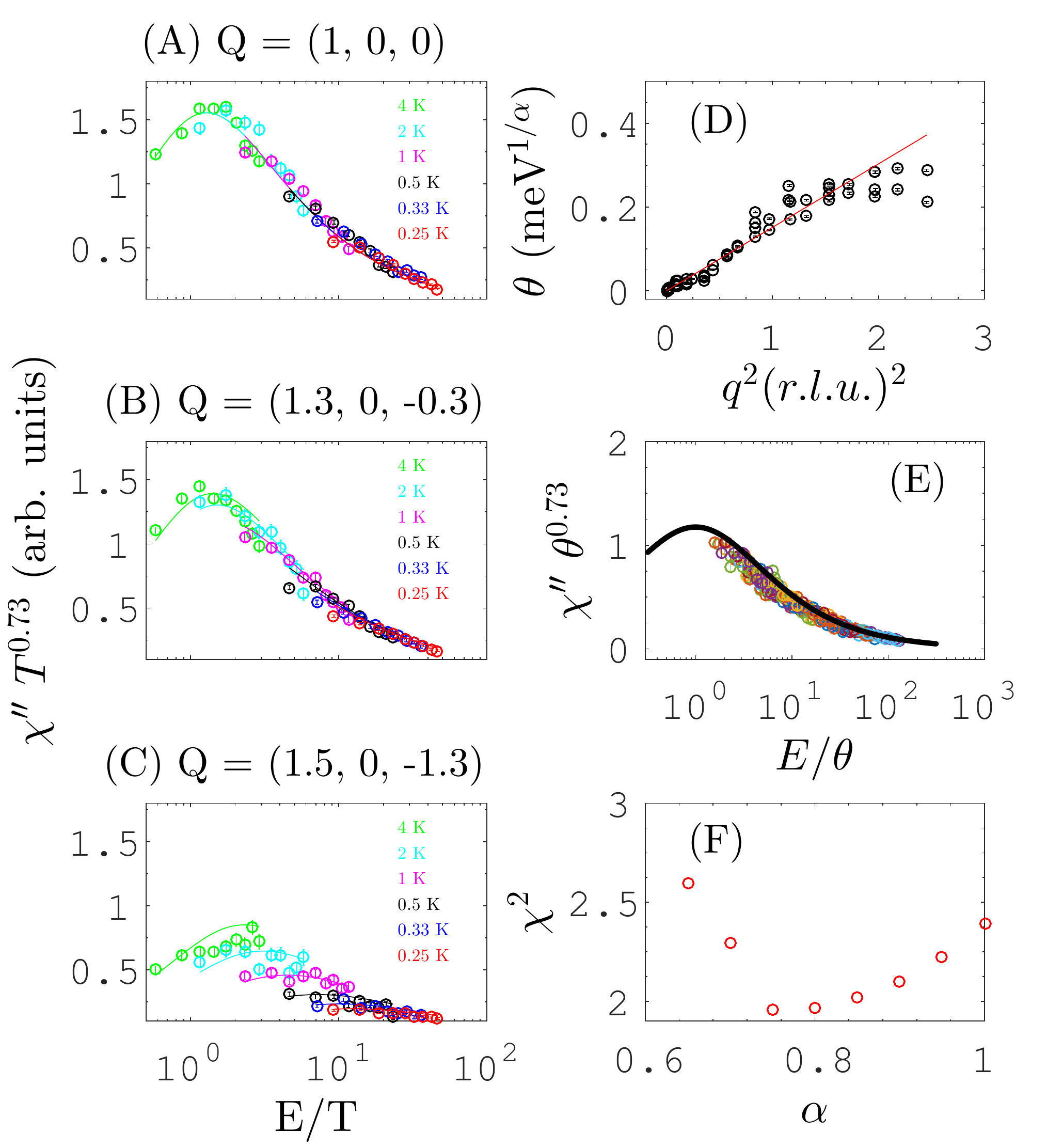}
\caption{Scaling analysis of \Agq. (A-C) $\chi^{\prime \prime} T^{\alpha}$ scaled as a function of $E/T$ in different regions of reciprocal space. The solid lines are a fit of eq. (\ref{scaling}) to the data. A global fit of eq. (\ref{scaling}) yields the scaling exponent $\alpha$=0.73(1). (D) $q$-dependence of $\theta$ determined from the fit shown in (A-C). Near $\bfq = \bfQ - \bfQ_0$ = 0, $\theta(q) \approx \bfq^2$ (red line). (E) $\chi"$ at 0.25 K scaled as a function of $E/\theta(q)$. $\chi"$ at different $\bfQ$s collapses onto a single curve showing $E/\theta(q)$-scaling. Each color represents a point in the (H 0 L) scattering plane. The solid line is a fit of eq. (\ref{scaling}) at zero $T$ to the data. (F) The variation in the goodness of fit parameter $\chi^2$ with $\alpha$.}
\label{pheno_plots}
\end{figure}

\begin{figure}
\includegraphics[width= .43\textwidth]{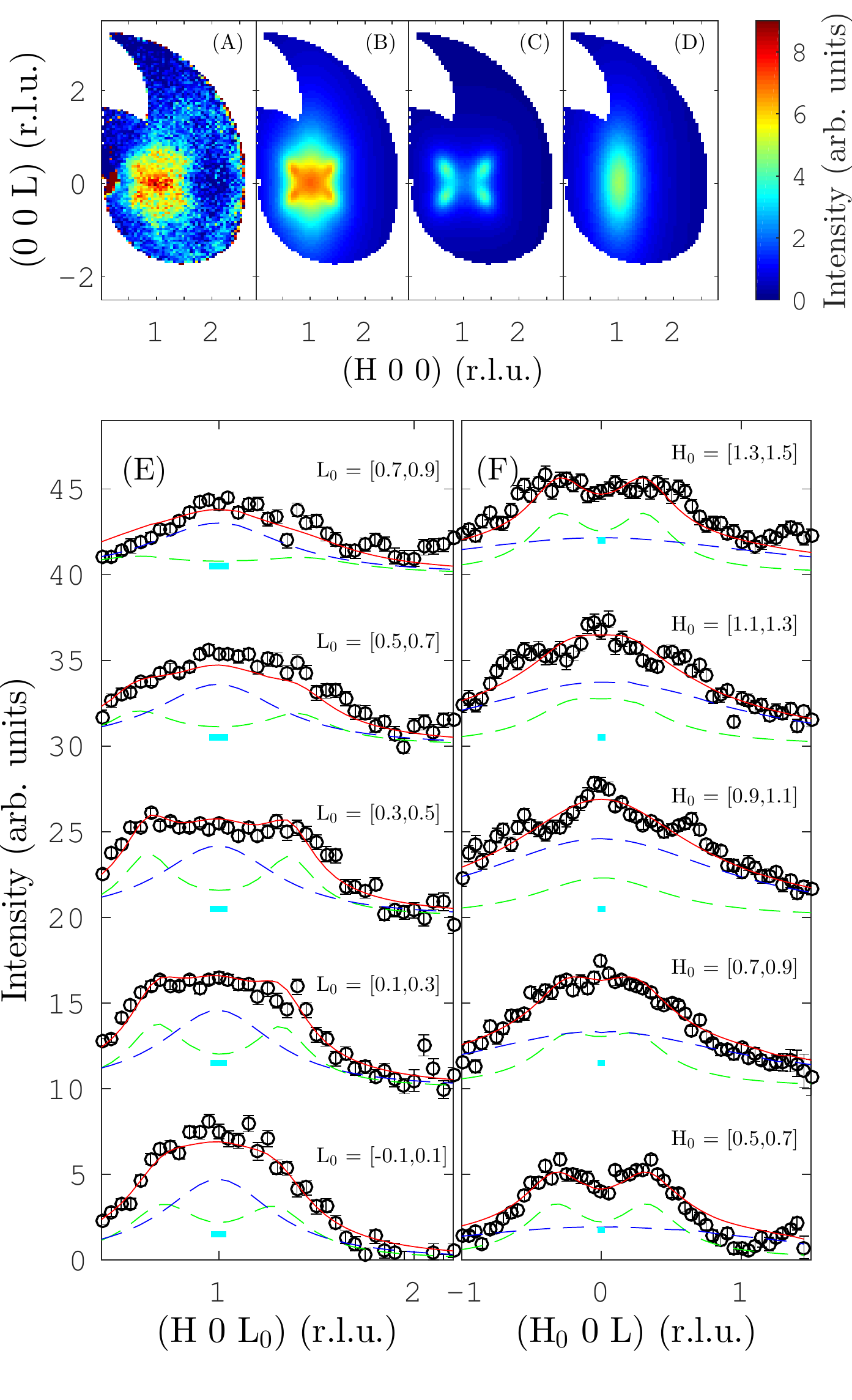}
\caption{ Parametrization of the diffuse magnetic scattering as a superposition of two independent magnetic fluctuations. (A) Constant energy slice at 0.25 K averaged over $E$ =[0.15,0.25] meV after subtraction of data collected at 50 K as a background. (B) The fitted pattern of two Lorentzians to the constant energy slice shown in (A). (C) and (D) show separately the two fluctuations ($\bfQ_1$ =(0.65 0 0.3) and $\bfQ_2$ = (1 0 0) respectively) contributing to (B). (E,F) Cuts along (E) (H 0 0) and (F) (0 0 L) with $E$ = [0.15,0.25] meV. The red line is the fit of two Lorentzian functions as described in the text. The green and blue dotted lines represent the contribution from the magnetic fluctuations centered at $\bfQ_1$ and $\bfQ_2$, respectively. Horizontal bars (cyan) indicate instrumental resolution. }
\label{cutscom}
\end{figure}

\begin{figure}
\centering
\includegraphics[width= 0.48\textwidth]{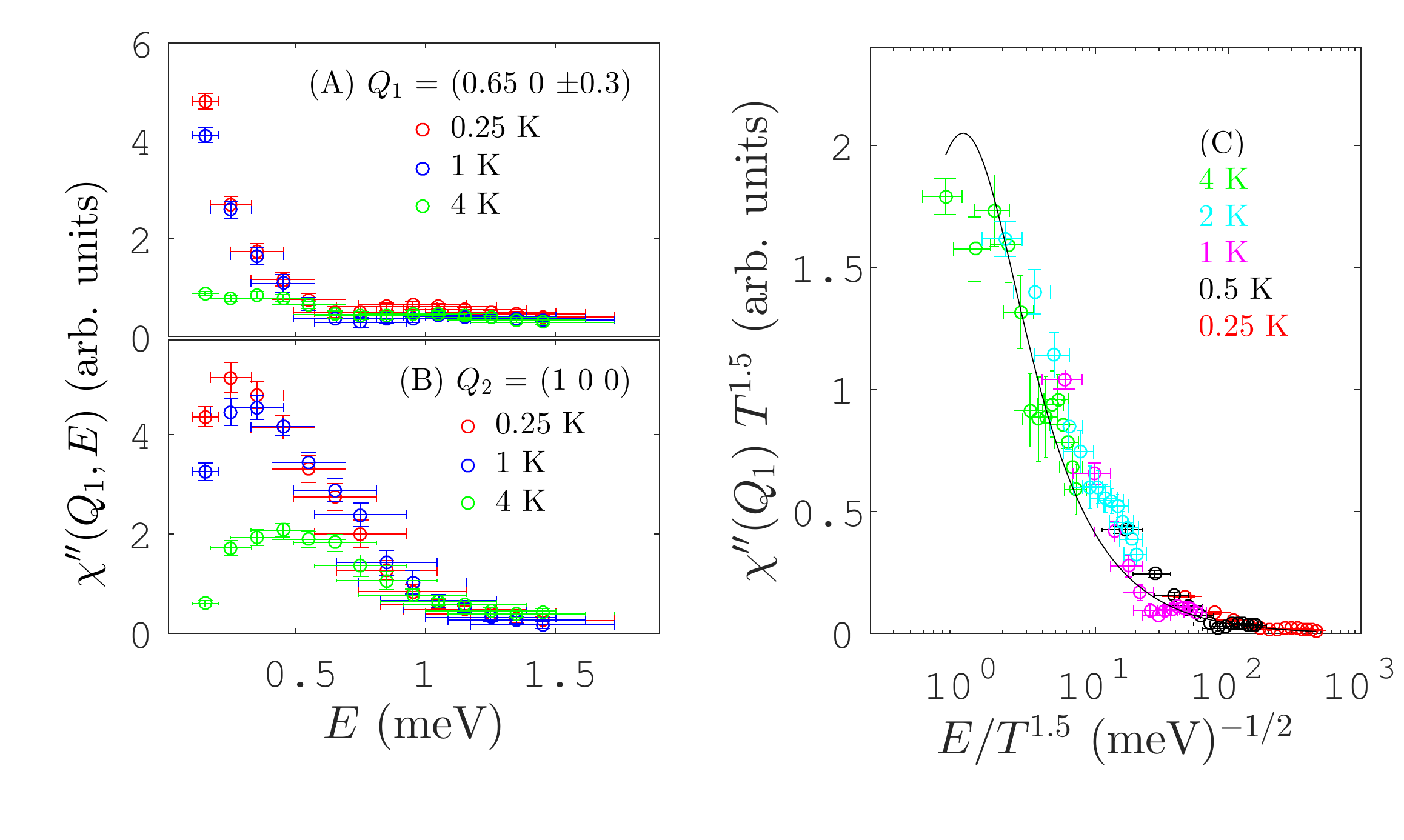}
\caption{Spectral weights of (A) $\bfQ_1$ = (0.65 0 $\pm$0.3) (B) $\bfQ_2$ = (1 0 0) extracted from the fit of 2D-Lorentzian function. (C) Scaling analysis of the component of $\chi"$ near $\bfQ_1$ using the HMM approach. This component of $\chi"$ is extracted from the fit of a sum of overlapping Lorentzians as shown in Fig. \ref{cutscom}. }
\label{separation}
\end{figure}

To make contact with the above picture, $\chi"$ was extracted from the INS measurements using the fluctuation dissipation theorem, $\chi^{\prime\prime} (\bfQ, E, T) = \pi(1-exp(-\frac{E}{k_BT}))S(\bfQ, E, T)$, where S(\bfQ, $E$, $T$) is the dynamic structure factor measured by INS. The quantity $\theta(\bfQ-\bfQ_0)$ was included as a fitting parameter without assumption of a particular functional form. A global fit of the imaginary part of equation (\ref{scaling}) was performed to $\chi"$ at various $\bfQ,T,E$. The fit yields $\alpha$ = 0.73(1) (see Fig. \ref{pheno_plots}(F) for $\alpha$ as a function of the goodness of fit). The measurement and fit for several regions of the (H 0 L) scattering plane are presented in Fig. \ref{pheno_plots}(a-c). For ease of comparison, $\chi^{\prime\prime}T^{\alpha}$ versus the dimensionless ratio $E/T$ is plotted. Consistent with eq. (\ref{scaling}), $\chi"$ at several temperatures collapses onto a single curve displaying $E/T$ scaling at $\bfQ_0$ = (1 0 0), which is the center of the diffuse structure (Fig. \ref{pheno_plots}(A)). The overlap becomes less pronounced as the magnitude of $\bfq = (\bfQ -\bfQ_0$) increases. The deviation from $E/T$-scaling at higher $\bfq$ is captured by the $\theta(\bfq)$ term of eq. (\ref{scaling}). The values of $\theta(\bfq)$ as a function of $q$ = |\bfQ -\bfQ$_0$| are presented in fig. \ref{pheno_plots}(D), which shows that $\theta(\bfq)$ varies quadratically near the center of the pattern (q = 0). As an alternative test, we performed $E/\theta(q)$ scaling at 0.25 K. As shown in Fig. \ref{pheno_plots}(E), $\chi"$ at 0.25 K is well fit by eq. (\ref{scaling}) and the quantity $\chi^{\prime\prime}(\theta(q))^{-\alpha}$ at different $\bfq$s collapses onto a single curve. The observed $E/T$ scaling with the exponent 0.73(1) implies that the QCP is unconventional. This observation is very similar to the related system \Aux \cite{schroder1998scaling,schroder2000onset}. The similarity of the $E/T$ scaling shown here, thermodynamic properties \cite{Lohneysen1996471,Scheidt2002133,Kuchler2003}, and microscopic details of the magnetic order \cite{poudel,Lohn_mag_order} demonstrate that the QCP in \Agx{} is virtually identical to the QCP in \Aux{} and thus the conclusions drawn here should have broad implications. 
\\

\hspace{-12pt}
\textbf{Multiple fluctuations.} A central question concerning the nature of the quantum critical behavior in \Agx~is: What type of spin-spin correlations in real space give rise to a butterfly shaped pattern and the observed $E/T$ scaling? The magnetic order in \Agx~is incommensurate with a wave-vector $\bfQ_1$ = (0.65 0 0.3) and is nearly independent of Ag composition \cite{poudel}. Hence, the critical scattering is expected at the wave-vector (0.65 0 0.3) and equivalent positions in reciprocal space, e.g. (1.35 0 0.3). However, fluctuations at only these spots are inconsistent with the maximum intensity occurring at $\bfQ_2$ = (1 0 0). This implies that there is at least one other competing fluctuation centered at the wave-vector $\bfQ_2$ = (1 0 0). The overlap of these two fluctuations renders the peculiar pattern of diffuse magnetic scattering. There is a natural relationship between these two types of fluctuations, with the fluctuation at (100) reflecting the tendency for commensurate rather than incommensurate order.

For a more comprehensive understanding of the critical fluctuations, the data in the (H 0 L) scattering plane was parameterized with a set of two-dimensional (2D) Lorentzian functions with the following form:
\begin{equation}
y =\frac{A}{1+\big(\xi_{\parallel}^2 \cos^2\theta+ \xi^2_{\perp} \sin^2\theta\big) \big[(Q_H - Q_{H0})^2 + (Q_L - Q_{L0})^2\big] }
 \label{2D_lorentz}
\end{equation}
Here, $\xi_{\parallel}$ and $\xi_{\perp}$ are the components of the real space correlation lengths parallel and perpendicular to an arbitrary axis. $\theta_0$ is the angle between the arbitrary axis and [H 0 0] in reciprocal space. Each Lorentzian is centered at a wave vector \bfQ{}= ($Q_{H0}, 0, Q_{L0}$). $\theta$ measures the angle between the wave-vector \bfq = ($Q_H -Q_{H0}$, 0, $Q_L - Q_{L0}$) and the arbitrary axis. A sketch showing the geometry in reciprocal space is shown in S5 \cite{supp}. Two sets of Lorentzian functions were used, one corresponding to $\bfQ_1$ and symmetry equivalent wave-vectors and the other to $\bfQ_2$. 

This parameterization provides a good description of the data for all temperatures and energy transfers and shows that the overlap of the fluctuations at $\bfQ_1$ and $\bfQ_2$ renders the peculiar pattern of diffuse magnetic scattering. An example at 0.25 K is shown in Fig. \ref{cutscom}(B), which consists of the 2D-Lorentzian components corresponding to $\bfQ_1$ and $\bfQ_2$ (Fig. \ref{cutscom}(C) and \ref{cutscom}(D) respectively). Cuts are also shown in Fig. \ref{cutscom}(E,F). The correlation length and the spectral weight of each fluctuation are estimated from a fit of eq. \ref{2D_lorentz} to the constant $E$ slices cut as a function of temperatures and energy transfer. The spectral weight of $\bfQ_1$ is shown in Fig. \ref{separation} (A), which shows that the fluctuation at $\bfQ_1$ is quasielastic. The correlation lengths along parallel and perpendicular axes are of the same order over the entire spectrum \cite{supp}. At 0.25 K and 0.2 meV, parallel, $\xi_{\parallel}$, and perpendicular, $\xi_{\perp}$, components of the correlation lengths of the fluctuation at $\bfQ_1$ are 47(2) $\mbox{\AA}$ and 69 (3) $\mbox{\AA}$, respectively. The correlation lengths being of the same order of magnitude indicates that the fluctuation at $\bfQ_1$ is clearly three dimensional. On the other hand, the fluctuation at $\bfQ_2$ appears to be inelastic with a small gap (Fig. \ref{separation}(B)). Similarly, the components of correlation lengths for the fluctuation at $\bfQ_2$ along the $a$ and $c$-axis are $\xi_a$=35(3) and $\xi_c$=9.3(5) $\mbox{\AA}$. 
\\

\hspace{-17pt}
\textbf{Critical Behavior.}
Taken together, the measurements show at least three distinct spin fluctuations in \Agq: $\bfQ_1$ = (0.65 0 0.3), $\bfQ_2$ = (1 0 0), and $\bfQ_3$ = (0 1 0). All three of these fluctuations occur on energy scales that influence the observed critical behavior. Understanding the critical behavior of this rich fluctuation spectrum is thus important. Here we focus on the critical behavior of the fluctuations $\bfQ_1$ and $\bfQ_2$, since these are most closely related to each other and to the magnetically ordered region of the phase diagram. There is a natural relationship between these two types of fluctuations: both fluctuations represent the tendency to form antiferromagnetic order with $\bfQ_1$($\bfQ_2$) reflecting the tendency for incommensurate (commensurate) order. The most straightforward expectation is that critical fluctuations should be found at $\bfQ_1$ as this corresponds to the wave vector of the antiferromagnetically ordered side of the QCP\cite{poudel}. The parameterization discussed above allows the energy dependence of scattering at $\bfQ_1$ and $\bfQ_2$ to be examined. The fluctuations centered at $\bfQ_1$ (Fig. \ref{separation}(A)) appear to be quasielastic, whereas the fluctuations centered at $\bfQ_2$ (\ref{separation}(B) appear to be inelastic with a small energy gap. This is in accord with the observed incommensurate antiferromagnetic ground state associated with $\bfQ_1$ as opposed to a commensurate antiferromagnetic ground state associated with $\bfQ_2$. It is interesting that the commensurate fluctuations at $\bfQ_2$ are observed in the parent compound CeCu$_6$ and thus may play an important role across a broad region of the phase diagram\cite{REGNAULT1987289,ROSSATMIGNOD1988376}.

Having determined that the magnetic scattering in \Agq~ comprises the critical fluctuations at $\bfQ_1$ in addition to the fluctuations at $\bfQ_2$ and $\bfQ_3$, we now analyze the critical part of $\chi"$ as a function of $E$ and $T$. For a 3D QCP, the HMM model predicts that $\chi"$ scales as a function of $E/T^{3/2}$ \cite{stockertfield,lohneysen2007magnetic}. When only the critical part of $\chi"$ is scaled as a function of energy and temperature, $E/T^{3/2}$-scaling is observed in \Agq. As shown in Fig. \ref{separation} (C), the quantity $\chi^{\prime\prime}T^{3/2}$ at different temperatures collapses onto a single curve which can be fit with the equation $\chi^{\prime\prime} =T^{-3/2}f(E/T^{3/2})$. This illustrates that at least a portion of the fluctuation spectrum of \Agx~is consistent with the HMM model.\\
\\
\hspace{-10pt}
\textbf{Discussion}\\
There are important implications of the above analysis. Firstly, while the $E/T$ scaling as well as the anomalous exponent of 0.73 is tantalizingly close to the expectations of several theoretical models \cite{Si_local,strong_coupling,strong_coupling_PNAS, dissi_xy,dissipation_Varma}, the critical magnetic fluctuations being 3D suggests that these models are inapplicable to the QCP studied here. Second, the E/T$^{3/2}$ scaling of the critical fluctuations indicates that long wavelength fluctuations of the order parameter play an important role, but the presence of other distinct spin fluctuations in the spectrum raises the prospect of coupling between order parameters.

One of the important puzzles in the study of QCPs is that field tuned QCPs in $\mathrm{CeCu_{6-x}}T_x$ ($T$ = Ag, Au) are characteristically different from the composition tuned QCPs\cite{Scheidt2002133,stockertfield,lohneysen2007magnetic}. The composition tuned QCPs in $\mathrm{CeCu_{6-x}}T_x$ are unconventional \cite{schroder2000onset,schroder1998scaling,Scheidt2002133}, while the field tuned QCPs in both systems appear to adhere to the framework of three dimensional HMM model \cite{stockertfield,lohneysen2007magnetic}. This dependence on tuning parameter is possible if the magnetic fluctuations undergo a  dimensional crossover (from 3D to 2D) approaching the composition tuned QCP. However, there is no reason at least from the structural perspective as to why magnetic fluctuations are two dimensional and such crossover occurs in $\mathrm{CeCu_{6-x}}T_x$. The results presented here provide a potential explanation to this puzzle. In the case of the field tuned QCP achieved by suppressing an antiferromagnetically ordered phase \cite{stockertfield,lohneysen2007magnetic}, the HMM scaling was found at the position of a magnetic ordering wave vector--which is an equivalent wave vector to where we find HMM scaling in \Agq{}. Thus the results here show that the nature of the QCPs are likely to be the same, and independent of whether the tuning mechanism is field or composition.  

The existence of competing magnetic fluctuations is a common feature of Ce-based QCPs. \Agx~nonetheless differs from most other Ce-based QCPs as the fluctuations are not well-separated in reciprocal space. A similar situation is also reported in the heavy fermion system $\mathrm{YbRh_2Si_2}$, in which ferromagnetic fluctuations compete with slightly incommensurate antiferromagnetic fluctuations and anomalous behavior such as a logarithmic divergence of $C/T$ and $E/T$ scaling of the dynamic susceptibility are observed near the QCP \cite{Cstock_YRS,Gegenwart2003} -- as in \Agx~and \Aux. Beyond these considerations, it is not entirely clear how the fluctuation spectrum results in $E/T$ scaling. The simplest explanation would be to take the separation of the fluctuation spectrum into two contributions at face value in which case the $E/T$ scaling would be the result of the interplay of critical and noncritical fluctuations. The nature of the coupling between the order parameters then becomes an important question \cite{morice2016hidden}. However, there are other interesting possibilities. For example, the fluctuation spectrum could be viewed as a collection of fluctuations to many different ordered states with only one ultimately going critical. An energy landscape corresponding to this scenario may explain the unusual thermodynamic properties observed for \Aux{} and \Agx{}. This might produce a similar situation to frustrated systems where $E/T$ scaling and a fractional exponent are observed \cite{spin_liquid}. An additional possibility is that there is a quantum Lifshitz point, although the available predictions of thermodynamic properties do not lend credence to this idea \cite{Ramazashvili}. Additional investigations as a function of composition and field will likely lead to deeper insight into the origin of $E/T$ scaling.

In conclusion, we performed extensive INS measurements of the critical fluctuations of the antiferromagnetic QCP in \Agq~as a function of $T$, $E$, and $\bfQ$. $\chi"$ near the QCP displays an $E/T$ scaling with an anomalous exponent of 0.73(1). Analysis over a large region of reciprocal space shows that there are at least three magnetic fluctuations with similar spectral weights. The portion of the fluctuation spectrum corresponding to the magnetically ordered region of the phase diagram scales as $E/T^{3/2}$ demonstrating that the critical behavior in \Agq~is consistent with the conventional three dimensional antiferromagnetic QCP, but more broadly the presence of other fluctuations suggests that coupling between order parameters plays in important role in the quantum critical behavior of the class of QCPs exemplified by \Agq.\\
\\
\hspace{-10pt}
\textbf{Methods}\\
\textbf{Material growth.} A single crystal of \Ag{} was grown using the Czochralski technique. Starting elements Ce (Ames Laboratory, purity = 99.998\%), Cu (Alpha Aesar, purity = 99.9999\%), Ag (Alpha Aesar, purity = 99.9999\%) were melted in a stoichiometric proportion. The Czochralski process was performed in a tri-arc furnace with a graphite hearth. The furnace was continuously purged with ultra high purity argon during the growth. The hearth was rotated with a constant speed of 100 revs/min. A seed rod was pulled with a constant vertical speed of 20 mm/hr. \\

\hspace{-10pt}
\textbf{Inelastic Neutron Scattering measurements:} Inelastic neutron scattering measurements in the (H 0 L) scattering plane of \Agq{} were carried out using the cold neutron chopper spectrometer (CNCS) of the Spallation Neutron Source (SNS) at Oak Ridge National Laboratory (ORNL). The measurement was performed with an incident energy ($E_i$) of 2.5 meV, which results in an elastic energy resolution of 0.07 meV. The crystal was rotated in the scattering plane, and for each position of the crystal, scattered neutrons were recorded by a large set of detectors that covers a horizontal angular range of $-50^{\circ}-+140^{\circ}$ and $\pm16^{\circ}$ in vertical direction. Additional measurements were performed in the (H K 0) scattering plane using the Multi-Analyzer Crystal Spectrometer (MACS) at the NIST center for neutron research (NCNR). The measurements with MACS was carried out with a fixed final energy ($E_f$) of 2.5 meV.


\vspace{10pt}
\hspace{-10pt}
\textbf{Acknowledgements}\\
We acknowledge W. Tian for help with sample characterization and C. D. Batista, T. Williams, R. Baumbach, J. W. Lynn, and N. P. Butch for useful discussions. The research at the Spallation Neutron Source at Oak Ridge National Laboratory is supported by the Scientific User Facilities Division, Office of Basic Energy Sciences, U.S. Department of Energy (DOE). AFM and DM acknowledge support from the U. S. DOE, Office of Science, Basic Energy Sciences, Materials Sciences and Engineering Division. Work at LANL was supported by the U.S. DOE, Basic Energy Sciences, Division of Materials Sciences and Engineering. The work at the NIST Center for Neutron Research utilized facilities supported in part by the National Science Foundation under Agreement No. DMR-1508249. This manuscript has been authored by UT-Battelle, LLC under Contract No. DE-AC05-00OR22725 with the U.S. Department of Energy. The United States Government retains and the publisher, by accepting the article for publication, acknowledges that the United States Government retains a non-exclusive, paid-up, irrevocable, world-wide license to publish or reproduce the published form of this manuscript, or allow others to do so, for United States Government purposes. The Department of Energy will provide public access to these results of federally sponsored research in accordance with the DOE Public Access Plan (http://energy.gov/downloads/doe-public-access-plan).\\

\hspace{-10pt}
\textbf{Author contributions}\\
L.P. and A.D.C. conceived the research project. L.P. and D.M. synthesized the single crystal for the measurement. A.F.M. and F.R. performed preliminary characterization of the sample. L.P., L.W., G.E., Y.Q. and A.D.C. performed the INS measurements. L.P. analyzed the data with help of J.M.L. and A.D.C. L.P., J.M.L., and A.D.C. prepared the manuscript with input from all authors.\\

\hspace{-10pt}
\textbf{Disclaimer}: The identification of any commercial product or a trade name does not necessarily imply endorsement or recommendation by the National Institute of Standards and Technology.

\newpage
\onecolumngrid
\pagebreak
\setcounter{equation}{0}
\setcounter{figure}{0}
\setcounter{table}{0}
\setcounter{page}{1}
\makeatletter
\renewcommand{\theequation}{S\arabic{equation}}
\renewcommand{\thefigure}{S\arabic{figure}}
\renewcommand{\bibnumfmt}[1]{[S#1]}
\renewcommand{\citenumfont}[1]{S#1}

\renewcommand{\vec}[1]{\bm{\mathbf{#1}}}
\newcommand{\mat}[1]{\bm{\mathbf{#1}}}
\newcommand{\trp}[1]{{#1}^{\intercal}}
\newcommand{\vhat}[1]{\vec{\hat{#1}}}
\newcommand{\tr}{\textrm{tr}}
\newcommand{\hc}{\textrm{h.c.}}
\newcommand{\h}[1]{{#1}^{\dagger}} 
\newcommand{\cc}[1]{{#1}^{*}}
\newcommand{\cb}[1]{\bar{#1}}

\begin{center}
\textbf{\large Supplemental materials for ``Unveiling the role of competing fluctuations at an unconventional quantum critical point''}
\end{center}
\vspace{0.25 in}
\twocolumngrid

\section*{Experimental details}

\subsection*{Characterization}

\begin{figure}[b!]
\centering
\includegraphics[clip, trim=0inch 0inch 0.0inch 0inch, width= 0.48\textwidth]{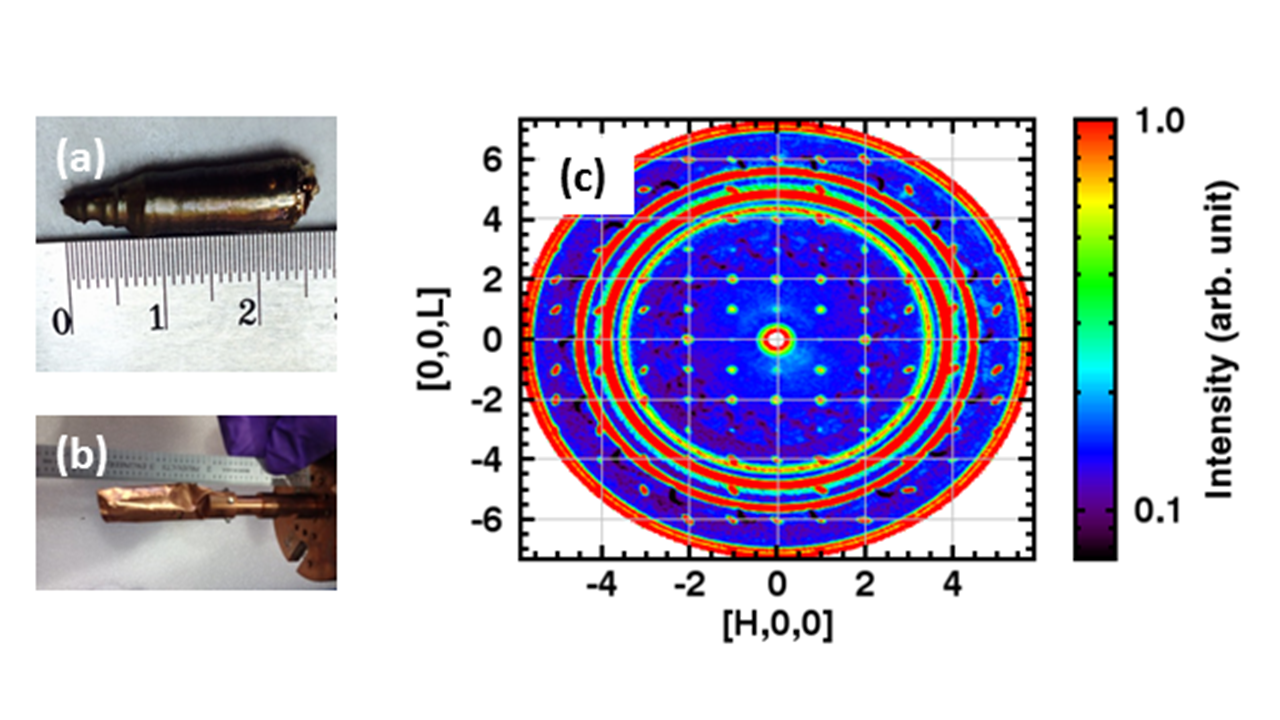}
\caption{(a) Single crystal of \Ag~ synthesized by the Czochralski process. (b) Single crystal mounted in a copper sample holder prior to the INS measurement. The sample is protected by a thin copper foil heat shield to ensure a uniform temperature in the sample. (c) Elastic neutron scattering measurement showing structural Bragg peaks in the (H 0 L) scattering plane. All major Bragg peaks are visible and can be indexed confirming that the sample is a large single crystal. }
\label{FIG1_supp}
\end{figure}

\begin{figure}[t]
\centering
\includegraphics[clip, trim=0.5inch 0inch 1inch 0inch, width= 0.485\textwidth]{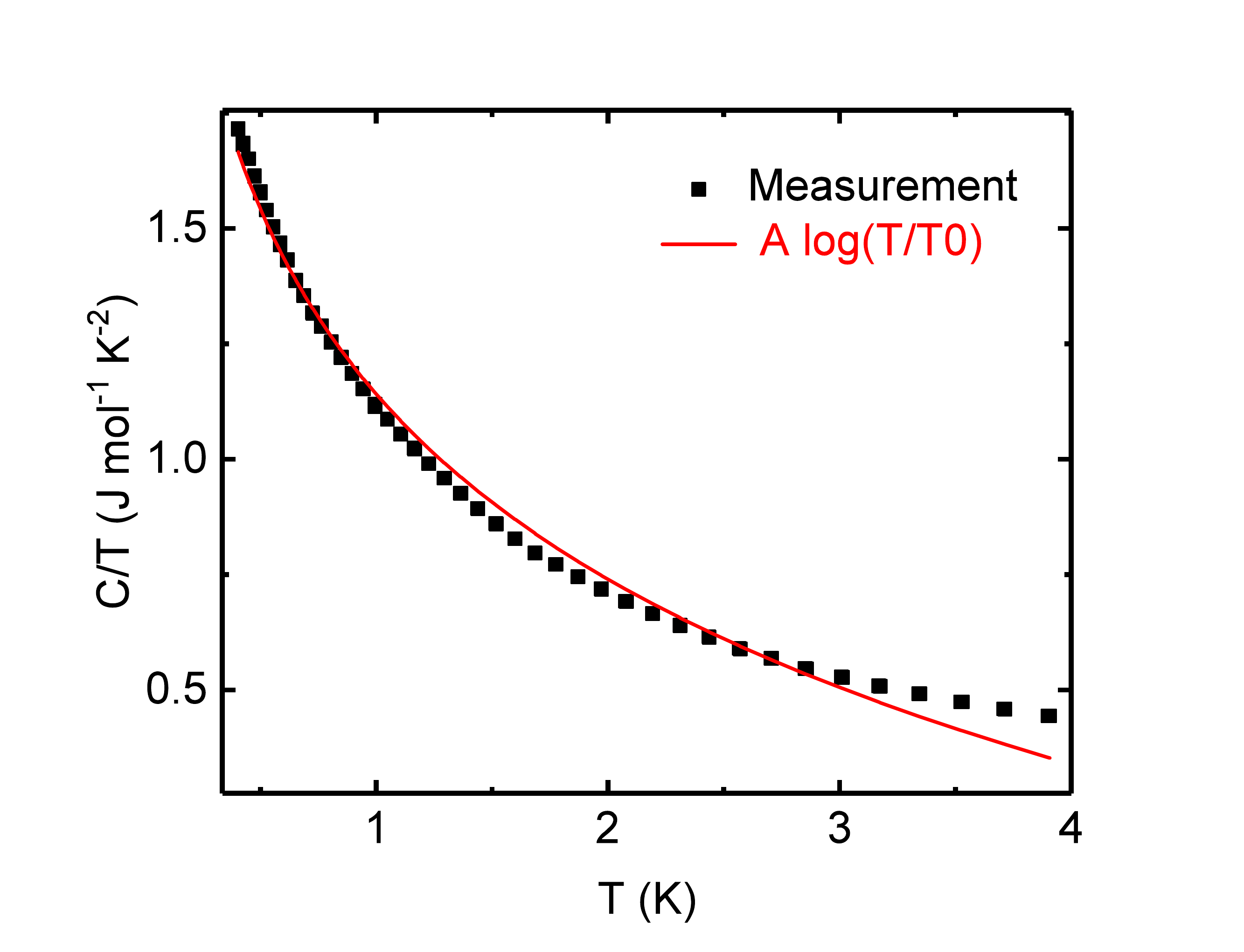}
\caption{Heat capacity with zero magnetic field displaying a divergence of $C/T$. The measurement is consistent with previous studies\cite{Kuchler2004}. The solid line is a fit to the functional form given in the legend which has been used previously to describe the heat capacity of CeCu$_{6-x}T_x$ ($T$ = Ag, Pd, Pt, Au) systems \cite{Kuchler2004,Sieck1996325}. }
\label{heat}
\end{figure}

Figure \ref{FIG1_supp}(a) shows the single crystal synthesized using the Czochralski process. The top and bottom part of the large single crystal were cut for energy dispersive x-ray spectroscopy (EDX) measurements, which indicates that both ends of the single crystal are homogeneous and that the crystal has no compositional variation. The crystal was aligned in the (H 0 L) scattering plane using the alignment station (CG-1b) at the high flux isotope reactor (HFIR) of Oak Ridge National Laboratory (ORNL). The sample was mounted in a copper holder and was covered by a thin copper foil to maintain a uniform sample temperature (Fig. \ref{FIG1_supp}(b)). A slice centered on the elastic line of the neutron scattering measurements shows well-indexed structural Bragg peaks attesting to the single crystallinity of the sample (Fig. \ref{FIG1_supp}(c)). Heat capacity (C) measurements were performed in a Quantum Design physical property measurement system (PPMS) with the $^3He$ option. Heat capacity at zero field shows a strong divergence of heat capacity over temperature ($C/T$) indicating that the sample is close to a QCP, Fig. \ref{heat}. The measurement is consistent with the literature\cite{Kuchler2004}. 

\begin{figure*}[t!]
\centering
\includegraphics[clip, trim=0inch 0inch 0inch 0inch, width= 0.85\textwidth]{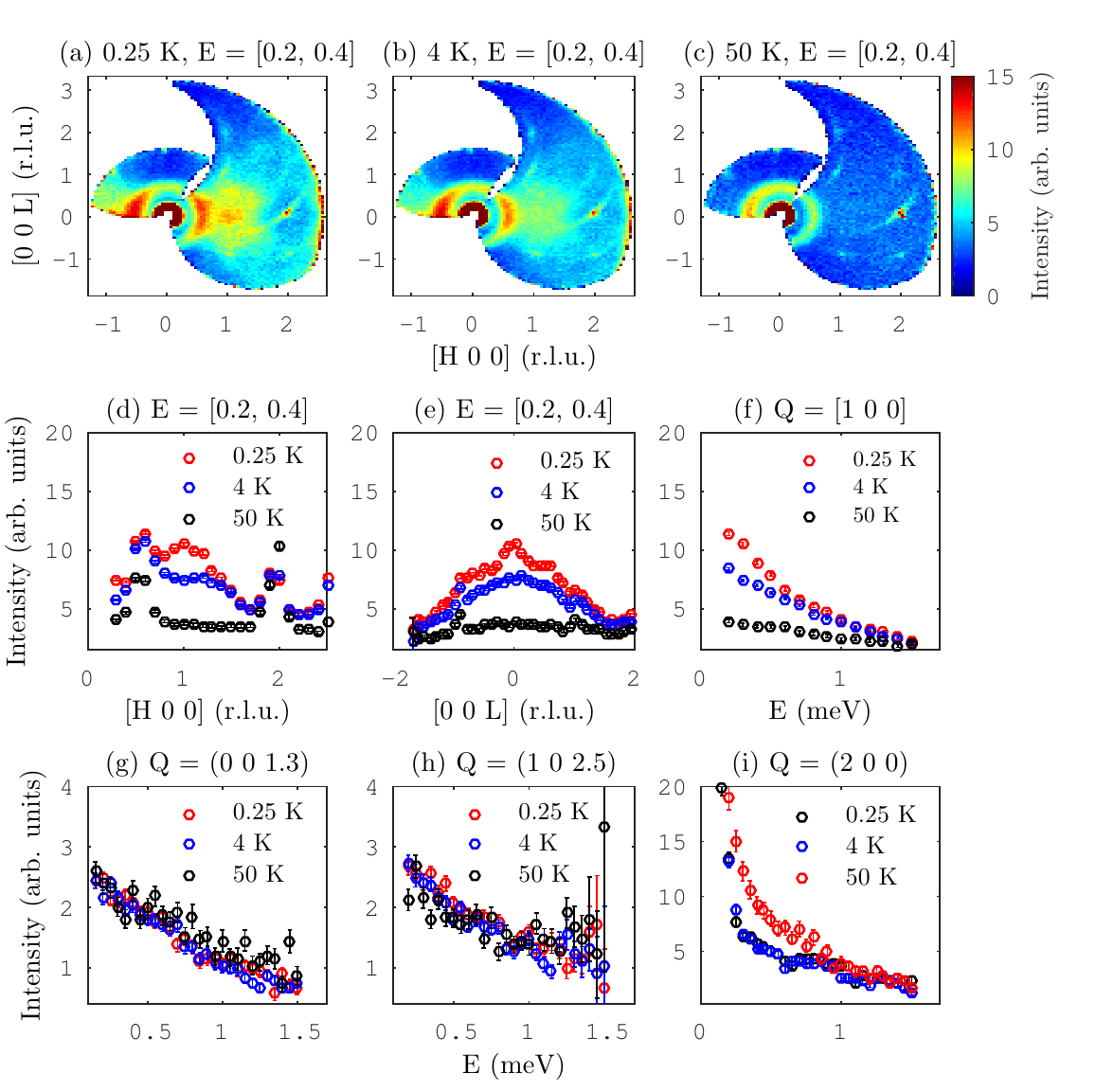}
\caption{(a,b,c) Constant energy slices at (a) 0.25 K (b) 4 K and (b) 50 K without a background correction. The slices are averaged over $E$ =[0.2,0.4] and $K$ =[-0.2,0.2]. The scattering observed at 50 K is primarily background from different sources. The scattering near \bfQ{}= (1 0 $\pm$ 1), \bfQ{}= (1 0 2), \bfQ{}= (2 0 $\pm$1) and \bfQ{}=(0 0 2) is due to the low frequency tail of the structural Bragg peaks. (d,e) Cuts at different temperatures along (d) [H 0 0] (e) [0 0 L] cut for the interval $E$ =[0.2,0.4] and $k$ =[-0.2,0.2]. (f) Cuts along $E$ near \bfQ{}= (1 0 0). (g,h) Cuts along $E$ at \bfQ{}= (0 0 1.3). The intensity at (0 0 1.3) and (1 0 2.5) does not vary with temperature. The \bfQ-independent background is taken from this wave-vector. (i) Energy cuts near the structural Brag peak \bfQ{}= (2 0 0). The intensity of the scattering increases at a high temperatures due the thermal population of phonons.}
\label{back_supp}
\end{figure*}

\begin{figure*}
\centering 
\includegraphics[clip, trim=0inch 0inch 0inch 0inch, width= 0.9\textwidth]{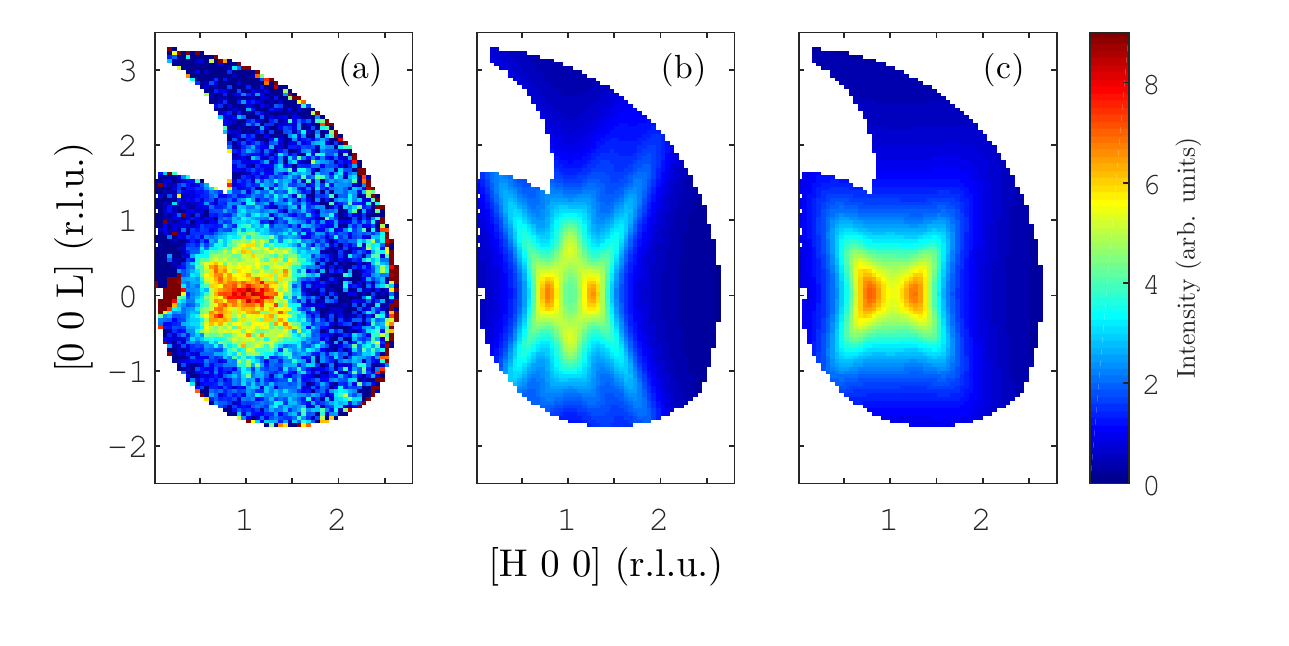}
\caption{(a) Constant $E$ slice at 0.25 K cut with $E$ = [0.075,0.275]. Background collected at 50 K is subtracted from the data. (b,c) The best fit of equation 2 of the main paper to the \bfQ{} slice obtained by (b) constraining the correlation lengths to differ by an order of magnitude, and (c) without constraints on the correlation lengths. }
\label{supp_rod_3D_model}
\end{figure*}

\begin{figure}[t]
\centering
\includegraphics[clip, trim=0inch 0inch 0inch 0inch, width= 0.48\textwidth]{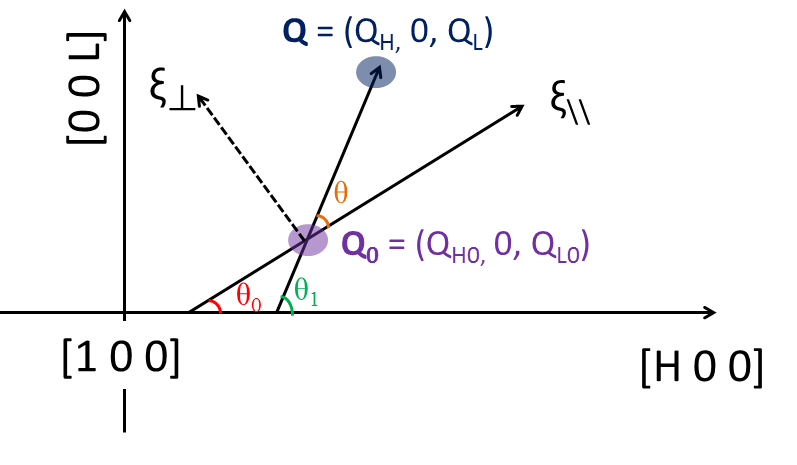}
\caption{Reciprocal space geometry showing the relative orientation of correlation length vector $\xi$ with \bfQ. $\xi_{\parallel}$ and $\xi_{\perp}$ are the parallel and perpendicular components of the correlation length. The parallel axis is at an angle $\theta_0$ from reciprocal space [H 0 0] axis (crystallographic $a$ axis in real space). Fitting the constant $E$ slice yields $\theta_0$ = 71$^\circ$. The magnetic scattering is centered at $\bfQ_0 = (Q_{H0},~0,~Q_{L0}$). $\bfQ{}= (Q_H,~0,~Q_L)$ is a wave vector in the scattering pane. 
 }
\label{diagram_geometry}
\end{figure}

\begin{figure*}[t]
\centering
\includegraphics[width= 0.75\textwidth]{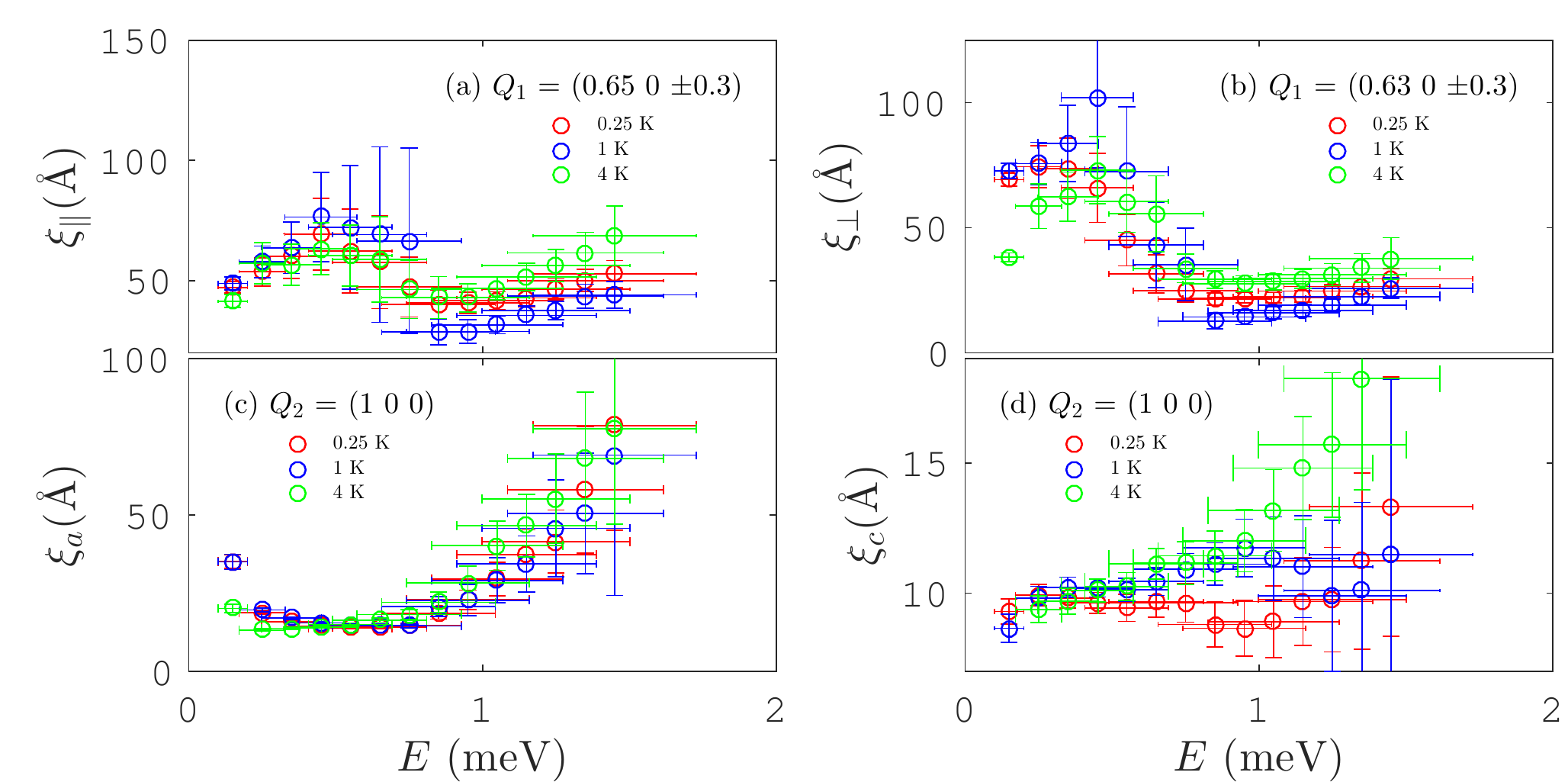}
\caption{Real space correlation lengths of the magnetic fluctuations. (a,b) Energy dependence of correlation lengths for the fluctuation centered at $\bfQ_1$ = (0.65 0 $\pm$0.3) along (a) parallel (b) perpendicular to the axis. (c,d) Energy dependence of correlation lengths for the fluctuation at $\bfQ_2$ = (1 0 0) along $a$ and $c$ axes. }
\label{correlation_lengths}
\end{figure*}

\subsection*{Neutron Scattering}
The measurement in the (H 0 L) scattering plane was carried out at eight different temperatures (0.25~K, ~0.325~K,~0.5~K,~1~K,~2~K,~4~K,~25~K,~50~K). For each temperature, the sample was rotated 110 degrees in the scattering plane in one degree steps. The data acquisition system at CNCS records time delay of each neutron in each detector's pixel for each angle measurement. The data in each pixel was converted into a four dimensional data set containing the dynamical correlation function $S(\bfQ, E)$, using the computer program Mantid \cite{mantid}. This reduction process considers the detector tubes efficiency, normalizes the data to total proton charge on target, and subtracts a time-independent background. This reduction was performed for each rotation angle. The set of $S(\bfQ,E)$ for all sample rotations were combined into a single data set using the Horace software suite \cite{horace}. 

The measurement in the (H K 0) scattering plane was carried out at two temperatures ($0.3~K,~50~K$) using MACS. For the measurement, the crystal and the detector were moved in a coupled way to map out all four quadrants of the scattering plane. Data analysis was performed using the computer program DAVE \cite{dave}. 

The data were sliced into two-dimensional color plots (e.g., see Fig. 1 of the main paper), or cut as one-dimensional plots along a selected $E$ or \bfQ{} axis.

\subsubsection*{Background Subtraction}
Constant $E$ slices, without background subtraction, with E= [0.2, 0.4] are shown in Figs. ~\ref{back_supp}(a,b,c). Plots are shown in Figs. ~\ref{back_supp}(d), \ref{back_supp}(e) and \ref{back_supp}(f), which represent cuts along H, L and E, respectively. The comparison of the slices and cuts shows that the intensity near \bfQ{}= (1 0 0) decreases with an increase in T, consistent with the assumed magnetic nature of scattering. The magnetic scattering disappears or becomes significantly weaker at 50 K. For all temperatures, a ring like structure is observed at low-Q due to the low angle instrumental background. The scattering along $\bfQ$ = [0 0 1.3], which is independent of $T$ and $E$ as shown in Fig. \ref{back_supp}(g)), is used to determine the $E$ dependence of the background. For instance, for the scaling analysis presented in Fig. 2 of the main paper, the non-magnetic background shown in the Fig. \ref{back_supp}(g) has been subtracted. As a check of consistency, we compare the background intensity at $\bfQ$ = (0 0 1.3) with $\bfQ$ = (1 0 2.5), which is shown in Fig. \ref{back_supp}(h)). The intensities being similar at two different regions of the scattering plane having different scattering angles supports the choice of the background. 

The \bfQ-dependence of the background can be obtained from the measurement at 50 K. As shown in Fig. \ref{back_supp}(c)), the intensity near (1 0 0) is similar to the intensity near \bfQ$_{BG}$ = (0 0 1.3), showing that the magnetic scattering is almost completely suppressed at 50 K. This comparison provides credence to the choice of the 50 K data as the background. Furthermore, we note that the 50 K measurements were performed with the same experimental configurations (sample rotation angle, cryostat, sample orientation, etc.) and account for instrumental background and contributions to the background that involve a scattering event in the sample. We note that the background near the (2 0 0) structural Bragg peak is affected by phonons. As shown in Fig. \ref{back_supp}(i), the intensity increases at a higher $T$ near the structural peak (2 0 0), which is attributed to the thermal population of low energy phonons due to the Bose factor. For the scaling analysis, the regions that are heavily affected by the phonons or the low-Q background are not used.

\section*{Dimensionality of the fluctuations}
 
Fig. \ref{diagram_geometry} provides the geometry in the reciprocal space showing the relative orientation of correlation length vectors with $\bfQ$. 

We compared the measurement with eq. 2 of the main paper under different assumptions (Fig. \ref{supp_rod_3D_model}). As is suggested in the literature \cite{stockert_2D}, we first made the assumption that the fluctuations are two-dimensional in real space, i.e., the correlation lengths along an arbitrary axis differ by an order of magnitude perpendicular to the axis. And since the critical wave-vector in \Agx~ is $\bfQ_1$ = (0.65 0 $\pm$0.3), the fluctuations are expected to be peaked at $\bfQ_1$ and related regions $\bfQ_1$ = (1.35 0 $\pm$0.3) of the scattering plane. A fit of four 2D- Lorentzians each centered either of (1.35 0 $\pm$0.3), (0.65 0 $\pm$0.3) was performed to the constant energy slice. To verify the assumption that the fluctuations are two-dimensional, the correlation lengths along the parallel and perpendicular axis are constrained to differ by at least an order of magnitude. The best fit obtained under that assumption is shown in Fig. \ref{supp_rod_3D_model}(b), which is clearly different from the measured slice shown in Fig. \ref{supp_rod_3D_model}(a). When the constraints on the correlation lengths are released, a slightly better fit is obtained (Fig. \ref{supp_rod_3D_model}(c)), although, the correlation lengths obtained from the fit are of the same order of magnitude, which rules out the possibility that the fluctuations are two-dimensional (2D). Therefore, the approach presented in the main paper, which considers an additional non-critical magnetic fluctuation is the simplest approach to describe the observed data. The scattering therefore can be parametrized as an overlap of two fluctuations near a critical wave-vector $\bfQ_1$ = (0.65 0 $\pm$0.3) and a non-critical term $\bfQ_2$ = (1 0 0). Under this assumption, the correlation lengths and the spectral weight of each fluctuation are estimated from a fit of two dimensional Lorentzian (eq. 2 of main paper) to the constant $E$ slices cut at different temperatures and energy transfers. The estimated correlation lengths for the fluctuations at $Q_1$ and $Q_2$ are shown in the Fig. \ref{correlation_lengths}.

\subsection*{Disclaimer}
The identification of any commercial product or a trade name does not necessarily imply endorsement or recommendation by the National Institute of Standards and Technology.

\end{document}